\documentclass[conference]{IEEEtran}
\usepackage[utf8]{inputenc}
\usepackage[english]{babel}
\IEEEoverridecommandlockouts
\usepackage{cite}
\usepackage{amsmath,amssymb,amsfonts}
\usepackage{algorithmic}
\usepackage{graphicx}
\usepackage{textcomp}
\usepackage{xcolor}
\def\BibTeX{{\rm B\kern-.05em{\sc i\kern-.025em b}\kern-.08em
    T\kern-.1667em\lower.7ex\hbox{E}\kern-.125emX}}

\usepackage[bookmarks=false]{hyperref}
\usepackage{url}
\usepackage{subcaption}
\newcommand{\superscript}[1]{\ensuremath{^{\textrm{#1}}}}

\def\affasterisk{\superscript{*}}
\def\affdag{\superscript{\dag}}
\def\affddag{\superscript{\ddag}}
\def\affS{\superscript{\S}}

\begin{document} 

\title{Interactive Discovery System for Direct Democracy\thanks{Accepted at the 2018 IEEE/ACM International Conference on Advances in Social Networks Analysis and Mining (ASONAM 18). For academic purposes, please cite the conference version.}}

\author{Pablo Aragón\affasterisk \affdag~~~~Yago Bermejo\affddag~~~~ Vicenç Gómez\affasterisk ~~~~ Andreas Kaltenbrunner\affasterisk \affS \\\\
  \begin{tabular}{ccccccc}
    {\affasterisk} Universitat Pompeu Fabra{\ } & & {\affdag} Eurecat{\ } & & {\affddag} Medialab Prado{\ } & & {\affS} NTENT{\ } \\
    Barcelona, Spain & & Barcelona, Spain & & Madrid, Spain & & Barcelona, Spain \\
  \end{tabular}\\\\
  \begin{tabular}{ccc}
    \{pablo.aragon,vicenc.gomez\}@upf.edu & yago@medialab-prado.es & kaltenbrunner@gmail.com \\
  \end{tabular}  
  
}


\maketitle


\begin{abstract}
Decide Madrid is the civic technology of Madrid City Council which allows users to create and support online petitions. Despite the initial success, the platform is encountering problems with the growth of petition signing because petitions are far from the minimum number of supporting votes they must gather. Previous analyses have suggested that this problem is produced by the interface: a paginated list of petitions which applies a non-optimal ranking algorithm. For this reason, we present an interactive system for the discovery of topics and petitions. This approach leads us to reflect on the usefulness of data visualization techniques to address relevant societal challenges.
\end{abstract}


\section{Introduction}
Madrid City Council launched in 2015 \textit{Decide Madrid}, a civic technology to foster open democracy and citizen participation through, among other features, online petitions\footnote{\url{https://decide.madrid.es/proposals}}. The functioning of this platform is as follows: First, a user publishes an online petition that proposes a public policy to be implemented in Madrid. Then, other users can comment the petition to favour deliberative practices of decision making, and users verified as citizens of Madrid can also support the petition. After one year, if the petition obtains more than 27,064 supporting votes (1\% of the population of Madrid over 16 years old), a voting process, advertised in the main page, is held. If most users vote in favour and the petition fulfills certain technical and ethical requirements, the proposal is implemented as a public policy by the City Council. In contrast, online petitions which do not reach the support threshold in one year are withdrawn from the platform.

In the first year of \textit{Decide Madrid}, two petitions reached the support threshold: one to unify public transport tickets\footnote{\url{https://decide.madrid.es/proposals/9-billete-unico-para-el-transporte-publico}} 
and the other to implement specific environmental sustainability policies\footnote{\url{https://decide.madrid.es/proposals/199-madrid-100-sostenible}}. 
Despite the success of both petitions, two aspects require special attention. First, both petitions were published on the day \textit{Decide Madrid} was launched. Second, no petition published in subsequent months has reached even half the threshold. According to previous analyses~\cite{betademic}, this might be explained by how petitions are presented in the platform. Originally, they were listed using pagination and a sorting criteria based on the number of supporting votes. As one could expect, this strategy embodied the so-called `Matthew effect'~\cite{Merton56}: the first petitions rapidly became popular and covered the first page of the ranking, which made new petitions nearly invisible. To overcome this effect, the City Council replaced the ranking algorithm with the \emph{Hot score} method of Reddit\footnote{\url{https://github.com/consul/consul/blob/master/lib/score_calculator.rb}}. However, this method was devised to rank news items whose interest usually declines rapidly, while online petitions require longer visibility to engage thousands of citizens and hence reach the threshold. Therefore, the results of this second strategy are still unsatisfactory and illustrate the socio-technical problem of ranking and filtering of information in social systems~\cite{ingo2014}.

Another problem of interest is the generalist approach of the interface, i.e., petitions are shown regardless of the user preferences.  Although some other ranking methods with personalized recommendation have been proposed~\cite{cantador2017personalized}, every approach has always relied on the assumption that petitions must be presented as a list and, therefore, requires a ranking algorithm. The system presented here is motivated by the use of alternative graphical interfaces to experiment with other strategies of human-computer interaction in \textit{Decide Madrid}. In particular, our system applies different techniques of data analysis and visualization to facilitate the discovery of topics and petitions\footnote{Code available at: \url{https://github.com/elaragon/decide-topics}}.

\begin{figure*}[!t]
\centering
\begin{subfigure}[b]{0.48\textwidth}
 \centering
 \includegraphics[trim={3cm 0cm 3cm 0cm},clip,width=0.99\textwidth]{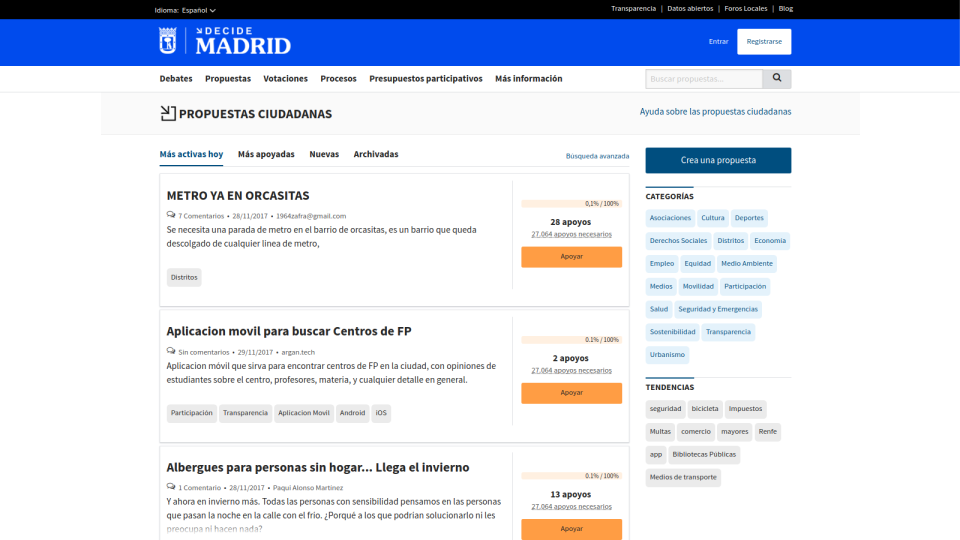}
\caption{\label{fig:fig1} Current front page of petitions in \textit{Decide Madrid}.}
 \end{subfigure}
 \hfill
\begin{subfigure}[b]{0.48\textwidth}
 \centering
 \includegraphics[trim={3cm 0cm 3cm 0cm},clip,width=0.99\textwidth]{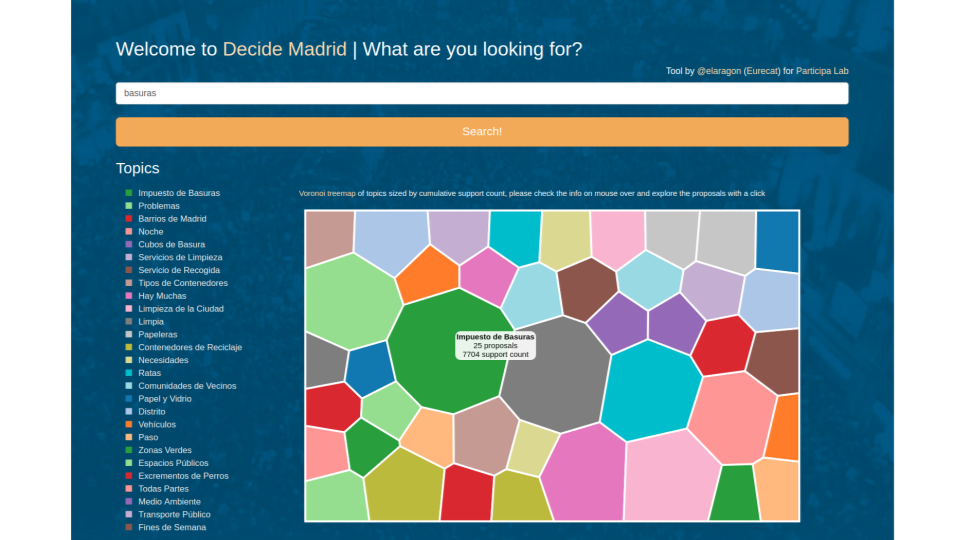}
\caption{\label{fig:fig2} Interface of topics in our interactive discovery system.}
 \end{subfigure}
 \\~\\8
\begin{subfigure}[b]{0.48\textwidth}
 \centering
 \includegraphics[trim={3cm 0cm 3cm 0cm},clip,width=0.99\textwidth]{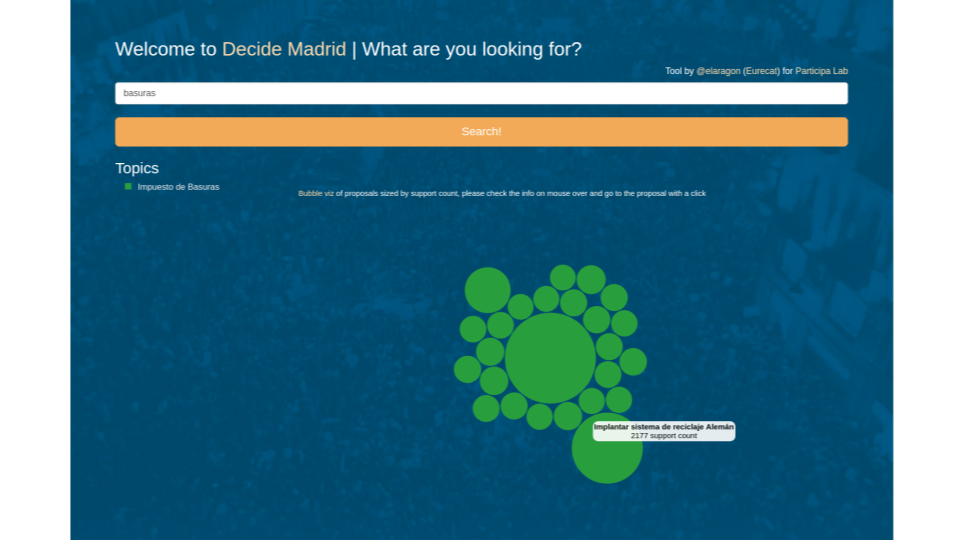}
\caption{\label{fig:fig3} Interface of petitions in our interactive discovery system.}
 \end{subfigure}
 \hfill
\begin{subfigure}[b]{0.48\textwidth}
 \centering
 \includegraphics[trim={3cm 0cm 3cm 0cm},clip,width=0.99\textwidth]{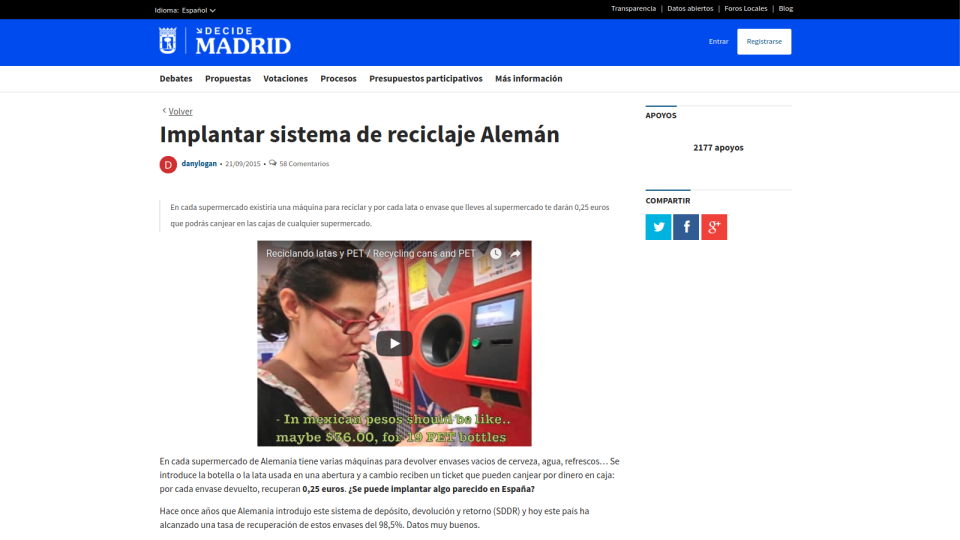}
\caption{\label{fig:fig4} Page of a petition in \textit{Decide Madrid}.}
 \end{subfigure}

\caption{Web interfaces of \textit{Decide Madrid} and the two modules of our interactive discovery system, using the query term ``basuras'', trash in Spanish. (a) Current front page of petitions in \textit{Decide Madrid} using a list of items sorted by the Hot Score, the default criteria. (b) Interface of the module of topics which groups online petitions into topics and shows them within a mosaic plot. (c) Interface of the second module which is loaded when a user clicks on a topic. The interface shows the corresponding petitions as circles sized by the number of supporting votes. (d) Page of a popular petition proposing the implementation of the German recycling system.} 
 \label{fig:screenshots}
\end{figure*}

\section{Related work}

The increasing abundance of data on urban activity has motivated the development of interactive data visualization systems (i.e., dashboards) in cities such as Madrid\footnote{\url{http://ceiboard.dit.upm.es/dashboard}}, Barcelona\footnote{\url{http://bcnnow.decodeproject.eu}}, 
Amsterdam\footnote{\url{http://citydashboard.waag.org}}, 
Dublin\footnote{\url{http://www.dublindashboard.ie}}, or Boston\footnote{\url{https://boston.opendatasoft.com}}. While dashboards of this type have typically relied on data from city sensors (e.g. traffic, noise, air pollution), open data portals, or social media platforms (e.g., Twitter, Facebook, Instagram), little effort has been devoted to deploy interactive discovery systems with data from civic participation platforms. One of the few cases is \emph{Civic CrowdAnalytics}, a tool which applies natural language processing with data from a crowdsourced urban planning process in Palo Alto~\cite{Aitamurto2016}. The tool served to reveal that the impact of citizens’ voices was related to the volume and the tone of their messages, i.e., demands with more messages and more emotive tone led to greater changes in public policies. Another example of interest is \emph{Pol.is}~\cite{polis2017}, a commenting and survey system. This tool applies principal component analysis to get major variances in opinions for different voters. These variances are then clustered using the k-means method, together with the silhouette score, to identify opinion groups. Although that platform was originally developed for news forums, it rapidly emph into a civic technology when adopted by the \textit{vTaiwan} community
to detect controversial issues about Uber/Airbnb Taiwanese regulations~\cite{polis}. 

For the specific case of \textit{Decide Madrid}, previous work focused on  interactive radial trees for the discussion threads of petitions~\cite{aragon2016ICWSMdemo}. Even though that work helped to provide a clear picture of the discussions, it assumed that users were able to find petitions of interest. Therefore, to overcome the socio-technical problems specified in the introduction, we describe below our novel interactive discovery system.

\section{System Design}
The interactive system is a pipeline of two modules which allow users to discover and explore two types of item: topics and petitions. To illustrate how the systems works, we will use an example based on petitions related to trash, a controversial issue in Madrid which has motivated many petitions on \textit{Decide Madrid} proposing different solutions to the existing problems.

\subsection{Interactive discovery of topics}

Figure~\ref{fig:fig1} shows the current web interface of \textit{Decide Madrid} which displays a list of petitions about any topic using the \emph{Hot score} method as sorting criteria by default. Although it allows users to retrieve petitions matching the term ``basuras'' (trash), they would have then to scroll through a dozen of pages to find proposals of interest among the 273 existing petitions. In contrast, the alternative web interface, shown in Figure~\ref{fig:fig2}, starts with a query form to force users to explicitly set their interest. Once a user has filled the form, the module of topics retrieves the matching petitions from the API\footnote{\url{https://decide.madrid.es/graphiql}} to then groups them into topics with the text clustering method Carrot\textsuperscript{2}~\cite{osinski2005carrot2}.
This is a state of art method for web search results clustering which is based on the the \emph{Suffix Tree Clustering} algorithm (STC)~\cite{Stefanowski:2003:CLP:1757448.1757479}. The idea behind this approach is that topics usually correspond to identical sequences of terms (phrases) and, therefore, groups of documents can be identified by such sequences. The algorithm follows two steps: 1) it first discovers groups of petitions with identical phrases, and 2) then merges these groups into larger clusters. Thus, the phrase which characterizes each final cluster is used by STC as a label to easily identify topics of petitions. Petitions can occur in multiple clusters and those without any assigned cluster can be found at ``Other topics''.

Resulting topics are presented in a mosaic plot with distinctive color for each topic and size according to the sum of supporting votes to the petitions of the corresponding cluster. Metadata of each topic (label and number of petitions and sum of supporting votes) are displayed in a tooltip on mouse over. Figure~\ref{fig:fig2} shows the topic visualization of the term ``basuras'' for which the discovery system automatically identifies 44 related topics, e.g., ``garbage bins'', ``garbage tax'', ``cleaning services''. The full list is presented in Table~\ref{tab:topics}. When the user clicks on a topic (either at the diagram or at the legend) the second module is loaded using the selected topic as input.

\subsection{Interactive discovery of petitions}

The web interface of the module of petitions is presented in Figure~\ref{fig:fig3}. To distinguish that items are now petitions of a selected topic, these are shown as circles. The radius of the circle is based on the number of supporting votes to easily identify which are the most popular petitions. However, to avoid a ranking of petitions using lists as done in \textit{Decide Madrid} so far, circles are displayed randomly without overlap by applying a force-directed graph layout~\cite{jakobsen2001advanced}.
Metadata of each petition (title and number supporting votes) are also shown in a tooltip on mouse over. The illustrative example in Figure~\ref{fig:fig3} corresponds to the petitions from the ``garbage tax'' topic. The cursor is over a popular petition which proposes the implementation of the German recycling system (\textit{Mehrwegpfand})\footnote{\url{https://decide.madrid.es/proposals/1824-implantar-sistema-de-reciclaje-aleman}}. When the user clicks on the circle, the petition web page is loaded in a new window (see Figure~\ref{fig:fig4}) to allow the user to review its proposals and to support it.

\section{Discussion}

In this paper we have presented an interactive system to visually browse and discover topics and petitions of \textit{Decide Madrid}. For the case study of petitions about trash, users would have had to explore the rankings of petitions page by page in the official web interface. Our system is a way to solve this task in a more dynamic manner. Because these paginated rankings of petitions have been identified as a barrier for petition signing, simple but effective data visualization can serve to prevent civic technologies from undesired effects of this kind. In recent years, many similar platforms have been developed to facilitate public participation and, ultimately, collective action~\cite{margetts2015political}. Nevertheless, research has provided evidence that social systems are very sensitive to how information is ranked and filtered~\cite{ingo2014}. Given the increasing popularity of civic technologies, our system is an illustrative example of how these platforms might benefit from presenting information in a concise and well-interpretable way. 

\section{Acknowledgments}
This work is supported by Medialab Prado and the Spanish Ministry of Economy and Competitiveness under the María de Maeztu Units of Excellence Programme (MDM-2015-0502). We would like to thank the team at Participa~LAB for their valuable feedback which served to improve this system.

\bibliographystyle{acm}
\bibliography{references}

\begin{table*}[]
\centering
\caption{Full list of discovered topics for the petitions retrieved from \textit{Decide Madrid} with the query term ``basuras'' (trash in Spanish). For each topic, the table includes an English traslation, the number of petitions, and the sum of supporting votes to the corresponding petitions. ``Other topics'' contains the petitions which were not assigned to any cluster by the \emph{Suffix Tree Clustering} algorithm.\\}
\label{tab:topics}
\scalebox{1.25}{

\begin{tabular}{l|l|r|r}
Topic                                    & English translation                & \begin{tabular}[c]{@{}l@{}}Number of \\ petitions\end{tabular} & \begin{tabular}[c]{@{}l@{}}Sum of supporting \\ votes to petitions\end{tabular} \\ \hline\hline
Noche                                    & Night                              & 32                                                             & 4,519                                                                            \\ 
Calles y Aceras                          & Streets and sidewalks              & 31                                                             & 4,256                                                                            \\ 
Caso                                     & Case                               & 30                                                             & 4,701                                                                            \\ 
Cubos de Basura                          & Garbage Bins                       & 30                                                             & 2,676                                                                            \\ 
Impuesto de Basuras                      & Garbage Tax                        & 30                                                             & 8,219                                                                            \\ 
Servicios de Limpieza                    & Cleaning Services                  & 28                                                             & 3,869                                                                            \\ 
Horarios de Recogida                     & Pickup Schedules                   & 26                                                             & 2,400                                                                            \\ 
También Hay                              & There are also                     & 25                                                             & 3,923                                                                            \\ 
Limpieza de la Ciudad                    & Cleaning the City                  & 23                                                             & 5,155                                                                            \\ 
Tipos de Contenedores                    & Types of Containers                & 23                                                             & 2,837                                                                            \\ 
Camiones de Basura                       & Garbage Trucks                     & 22                                                             & 1,433                                                                            \\ 
Suciedad                                 & Dirt                               & 22                                                             & 5,236                                                                            \\ 
Comunidades de Vecinos                   & Neighborhood Communities           & 21                                                             & 2,630                                                                            \\ 
Papeleras                                & Litter bins                        & 21                                                             & 4,665                                                                            \\ 
Centro de Madrid                         & Centre of Madrid                   & 20                                                             & 2,798                                                                            \\ 
Contenedores de Reciclaje                & Recycling Containers               & 19                                                             & 3,234                                                                            \\ 
Mucha Gente                              & Many People                        & 19                                                             & 4,383                                                                            \\ 
Mucho Menos                              & Much Less                          & 15                                                             & 2,038                                                                            \\ 
Zonas Verdes                             & Green Zones                        & 14                                                             & 2,005                                                                            \\ 
Excrementos de Perros                    & Dog Droppings                      & 11                                                             & 1,637                                                                            \\ 
Malos Olores                             & Bad Smells                         & 10                                                             & 1,179                                                                            \\ 
Medio Ambiente                           & Environment                        & 10                                                             & 1,599                                                                            \\ 
Barrio Limpio                            & Clean Neighborhood                 & 9                                                              & 760                                                                             \\ 
Transporte Público                       & Public Transportation              & 9                                                              & 1220                                                                            \\ 
Vía Pública                              & Public Roads                       & 9                                                              & 918                                                                             \\ 
Plazas de Aparcamiento                   & Parking Spaces                     & 8                                                              & 368                                                                             \\ 
Calidad de Vida                          & Quality of Life                    & 7                                                              & 1,492                                                                            \\ 
Descampados                              & Open fields                        & 7                                                              & 605                                                                             \\ 
Solares Abandonados                      & Abandoned Plots                    & 7                                                              & 1,735                                                                            \\ 
Cantidad de Dinero                       & Amount of Money                    & 6                                                              & 3,548                                                                            \\ 
Parques Infantiles                       & Children's playgrounds             & 6                                                              & 362                                                                             \\ 
Residuos Orgánicos                       & Organic Waste                      & 6                                                              & 307                                                                             \\ 
Vehículos sin Motor                      & Non-Motor Vehicles                 & 6                                                              & 325                                                                             \\ 
Nuestros Hijos                           & Our Children                       & 5                                                              & 209                                                                             \\ 
Cualquier Sitio                          & Any Site                           & 4                                                              & 552                                                                             \\ 
Problemas de Movilidad                   & Mobility Problems                  & 4                                                              & 658                                                                             \\ 
Teniendo en Cuenta                       & Taking into account                & 4                                                              & 186                                                                             \\ 
Debería Hacerse                          & It Should Be Done                  & 3                                                              & 573                                                                             \\ 
Distrito más Poblado                     & Most Populous District             & 3                                                              & 340                                                                             \\ 
Control de Plagas                        & Pest Control                       & 2                                                              & 291                                                                             \\ 
Necesidad de Buscar                      & Need to Search                     & 2                                                              & 81                                                                              \\ 
Pasar por Caja el Producto               & Checkout the Product               & 2                                                              & 2,256                                                                            \\ 
Redes Sociales                           & Social Networks                    & 2                                                              & 250                                                                             \\ 
Vivienda y te Combiertas en un Proscrito & Housing and Commute into an Outlaw & 2                                                              & 3,183                                                                            \\ 
Other Topics                             & Other Topics                       & 60                                                             & 9,388                                                                            \\ 
\end{tabular}

}
\end{table*}

\end{document}